\documentclass[journal]{IEEEtran}

\usepackage{amsmath,amsfonts,amssymb,amsthm}
\usepackage{algorithmic}
\usepackage{algorithm}
\usepackage{array}
\usepackage{graphicx}
\usepackage{subcaption}
\usepackage{caption}
\usepackage{textcomp}
\usepackage{stfloats}
\usepackage{url}
\usepackage{verbatim}
\usepackage{color} 
\graphicspath{{./figs/}}
\usepackage[square,numbers, sort&compress]{natbib}

\usepackage[hidelinks]{hyperref}
\usepackage{booktabs}
\usepackage[capitalise]{cleveref}
\crefname{equation}{}{}
\usepackage[most]{tcolorbox}
\usepackage{multicol}
\usepackage{pifont}
\usepackage{threeparttable}
\usepackage{acro}
\DeclareAcronym{CRLB}{
  short = CRLB,
  long = Cramér-Rao lower bound,
}

\DeclareAcronym{EKF}{
  short = EKF,
  long  = extended Kalman filter,
}

\DeclareAcronym{FGO}{
    short = FGO,
    long = factor-graph-based optimization,
}

\DeclareAcronym{GNSS}{
  short = GNSS,
  long = global navigation satellite system,
}

\DeclareAcronym{GP}{
  short = GP,
  long  = Gaussian process,
}

\DeclareAcronym{INS}{
  short = INS,
  long  = inertial navigation system,
}

\DeclareAcronym{PF}{
  short = PF,
  long  = particle filter,
}

\DeclareAcronym{KF}{
  short = KF,
  long  = Kalman filter,
}

\DeclareAcronym{RLS}{
  short = RLS,
  long = recursive least squares
}
\DeclareAcronym{RBPF}{
  short = RBPF,
  long = Rao-Blackwellized particle filter,
}
\DeclareAcronym{SLAM}{
  short = SLAM,
  long  = simultaneous localization and mapping, 
}

\DeclareAcronym{UKF}{
  short = UKF,
  long  = unscented Kalman filter,
}

\DeclareAcronym{SQUID}{
  short = SQUID,
  long  = superconducting quantum interference device,
}
\usepackage{mathtools}
\usepackage{xspace}
\usepackage{tabularx}
\usepackage{array}
\usepackage{booktabs}

\usepackage{pgfplots}
\usepgfplotslibrary{fillbetween}
\usepackage{xcolor}
\pgfplotsset{compat=1.18}

\usepackage{tikz}
\tikzset{
    dashdot/.style={
        dash pattern=on 3pt off 3pt on 1pt off 3pt
    }
}
\usepackage{pgfplots}
\usetikzlibrary{arrows.meta,calc,decorations.pathreplacing,positioning,shapes,arrows,calc}

\newcommand\eg{e.g.\@\xspace}

\newcommand{\Transp}{\mathsf{T}}

\newcommand{\M}{\ensuremath{\mathcal{M}}\xspace}
\newcommand{\B}{\ensuremath{\mathcal{B}}\xspace}

\begin{document}
\title{Magnetic-Field-Based Localization Using Spatial Field Variations: Signal Processing Principles, Models, and Challenges}

\author{Isaac~Skog,~\IEEEmembership{Senior~Member,~IEEE},
Manon~Kok,~\IEEEmembership{Senior~Member,~IEEE},
Christophe~Prieur,~\IEEEmembership{Fellow,~IEEE},
and
Gustaf~Hendeby,~\IEEEmembership{Senior~Member,~IEEE}
\thanks{I.~Skog is with the KTH Royal Institute of Technology, Sweden, FOI Swedish Defence Research Agency, and Digital Futures. email: \emph{skog@kth.se}}
\thanks{M.~Kok is with the Delft University of Technology, the Netherlands. email: \emph{m.kok-1@tudelft.nl}}
\thanks{C.~Prieur is with the University Grenoble Alpes, CNRS, Grenoble-INP, GIPSA-Lab, F-38000 Grenoble, France. email: \emph{christophe.prieur@gipsa-lab.fr}}%
\thanks{G.~Hendeby is with Linköping University, Sweden. email: \emph{gustaf.hendeby@liu.se}}%
\thanks{Manuscript received ...}%
}

\markboth{IEEE Signal Processing Magazine,~Vol.~XX, No.~XX, Dec~2025}%
{\MakeLowercase{\textit{et al.}}: Author Guidelines for Special Issue Articles of IEEE SPM}

\maketitle

\IEEEPARstart{S}{ignal} processing has played, and continues to play, a fundamental role in the evolution of modern localization technologies. Localization using spatial variations, a.k.a. anomalies, in the Earth's magnetic field is no exception. It relies on signal-processing methods for statistical inference, magnetic-field modeling, and sensor calibration. Contemporary localization techniques based upon spatial variations in the magnetic field can provide decimeter-level localization accuracy indoors~\cite{liGDR:2012}, and outdoor localization accuracy on par with strategic-grade inertial navigation systems~\cite{muradoglu2025quantum}. This article provides a broad, high-level overview of current signal-processing principles, models, and open research challenges in localization using spatial variations in the Earth's magnetic field. The aim is to provide the reader with an understanding of the similarities and differences among existing key technologies, from a statistical signal-processing perspective. To that end, these existing key technologies will be presented within a common parametric signal-model framework compatible with well-established statistical inference methods. A comprehensive treatment of all types of magnetic-field-based localization technologies is beyond the scope of this article; interested readers are referred to the survey articles~\cite{Ouyang2022,Pasku2017,Lei2025,He2017}, which have different focuses. Specifically,~\cite{Ouyang2022} describes existing techniques for magnetic-field indoor localization, and challenges related to large-scale use of these techniques,~\cite{Pasku2017} provides a high-level system description of magnetic-field localization, including the generation of artificial magnetic fields,~\cite{Lei2025} focuses on robotics applications, and~\cite{He2017} focuses on smartphone-based indoor localization.

The use of the Earth’s magnetic field for localization and navigation dates back over a millennium to the introduction of the first compasses. However, it was not until about half a century ago that research on the use of spatial variations in the Earth's magnetic field for localization began~\cite{Goldenberg2006}. The main application areas were then aerospace and marine systems. With the introduction of miniaturized electronic magnetometers in the early 2000s and their subsequent integration into smartphones, the field of indoor localization for robots and humans using magnetic-field variations flourished. Here, the signal-processing community has played, and continues to play, a crucial role by providing theoretically sound and efficient principles for modeling sensor measurements~\cite{sieblerLSH:2023} and the spatial variation in the magnetic field~\cite {angermannFDJR:2012}, as well as for statistical inference of location and model parameters~\cite{kokS:2018}.

Recently, driven by geopolitical concerns about jamming and spoofing of \acp{GNSS} and other radio-based localization technologies, there has been renewed interest in magnetic-field localization techniques for aerospace and marine systems, as well as various emerging autonomous systems. Magnetic field-based localization is passive and thereby stealthy, and challenging to jam on a large scale because the field strength decays cubically with distance in the far field. This renewed interest in using magnetic-field-based localization techniques as a complement to \ac{GNSS}-based localization, along with the emergence of affordable, power-efficient, and compact high-performance magnetometers based on quantum sensing technology, has created new research challenges and opportunities. 

The outline of the article is as follows. First, key technologies for localization based on spatial variations in the magnetic field are introduced from a Bayesian perspective. Subsequently, mathematical models of magnetic fields and magnetometers are introduced and used to define a parametric signal model framework. Thereafter, statistical inference for magnetic-field map learning and localization using the parametric signal model framework is discussed. The article concludes with a discussion on open research challenges and a technology outlook.

\section*{Introduction}\label{sec:introduction}
\begin{figure*}[tb!]
    \centering
    \begin{subfigure}[t]{0.48\textwidth}
        \centering
        \includegraphics[width=\linewidth]{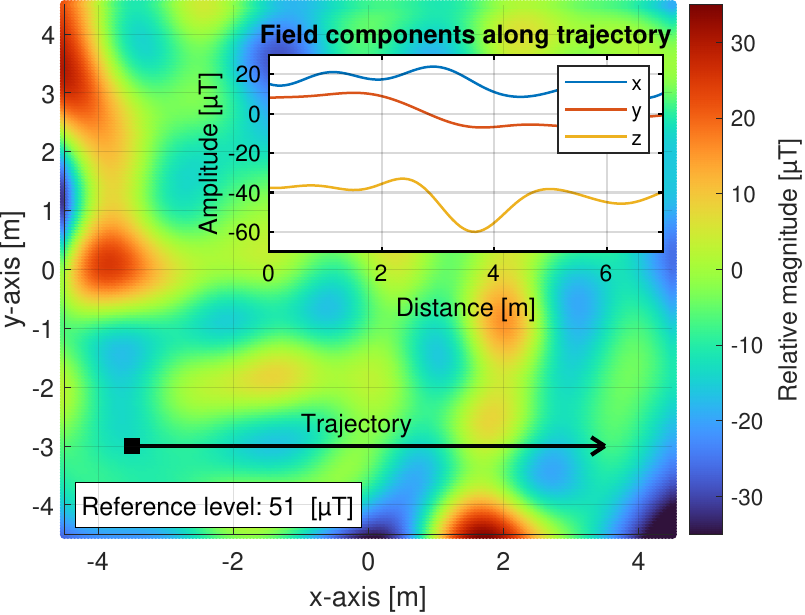}
        \caption{Indoor magnetic-field magnitude variations.}
        \label{fig:visionen_field}
    \end{subfigure}
    \hfill
    \begin{subfigure}[t]{0.48\textwidth}
        \centering
        \includegraphics[width=\linewidth]{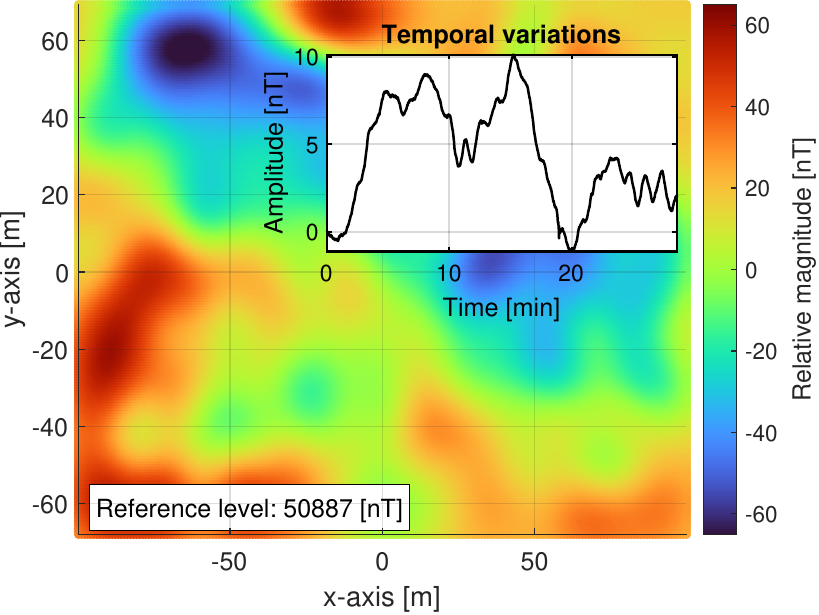}
        \caption{Outdoor magnetic-field magnitude variations.}
        \label{fig:granso_field}
    \end{subfigure}
    \caption{Examples of the spatial variations in the magnetic-field magnitude indoors and outdoors. Also shown in (\subref{fig:visionen_field}) are the three field components along the trajectory indicated by the black arrow. Also shown in (\subref{fig:granso_field}) is an example of temporal variations in the magnetic field during 30 minutes. Note the significant difference in length and magnitude scale indoors and outdoors.}
    \label{fig:fields}
\end{figure*}

Although the Earth's magnetic field is commonly approximated as a constant dipole field, local spatial variations in the ambient magnetic field are omnipresent. The spatial variations are caused by ferromagnetic materials in the Earth's crust and by man-made structures, which interact with the magnetic field generated by the Earth’s core~\cite{lowrie2023earth,canciani2016absolute}. Two examples of these field variations, indoor and outdoor, are shown in Fig.~\ref{fig:fields}. Although the figure illustrates the spatial variations in field magnitude, it is worth emphasizing that the magnetic field is a three-component vector field, and these spatial variations occur in all three components; see Fig.~\ref{fig:visionen_field}. Moreover, the field is also temporally varying due to magnetospheric disturbances, solar storms, etc.; see Fig.~\ref{fig:granso_field}. These temporal variations are typically on the order of tens to hundreds of nano-Tesla. For indoor localization, where the spatial variations are on the order of tens of micro-Tesla, they are typically neglected. In outdoor localization, where spatial variations are on the order of tens to hundreds of nano-Tesla, they may be compensated for using data from a stationary reference magnetometer or using filtering~\cite{canciani2016absolute}. Since this article focuses on localization using spatial variations in the magnetic field, temporal variations are disregarded unless explicitly stated otherwise, and all fields are assumed static.

\subsection*{Localization techniques}
Localization refers to the process of inferring the state $x_t$ of a localization system at time instant $t$  from measurements. The state $x_t$ at least includes the location $r_t$, but is commonly augmented with quantities, such as the orientation $q_t$ and velocity $v_t$, required to describe the sensor measurements and the motion of the platform carrying the localization system. 

There are three main techniques for localization using spatial variations in the Earth's magnetic field $\B$: map matching, \ac{SLAM}, and dead reckoning.

\emph{Magnetic-field map matching}, where the state $x_t$ is inferred by matching a sequence of magnetometer measurements $y_{1:t}$ to a map~\M. The map, directly or indirectly, describes the magnetic field~\B, the magnitude $\lVert \B\rVert$, or the field potential as a function of the location $r_t$. If the features within the map~\M are linked to a global frame of reference, this technique provides absolute location with a bounded location error.  

\emph{Magnetic-field \ac{SLAM}}, where the magnetometer measurements $y_{1:t}$ are simultaneously used to construct a map \M and infer the state $x_t$ with respect to this map. Without additional measurements linked to a global frame of reference, this technique only provides relative location with respect to the initial system state $x_0$. When new areas are mapped, the location error grows unboundedly, whereas revisiting already mapped areas keeps the error bounded.
  
\emph{Magnetic-field dead reckoning}, where information about translation and orientation changes is extracted from the magnetic field using a low-complexity time-varying map, and accumulated to infer the state $x_t$ relative to an initial pose $x_0$. This can be viewed as a degenerate form of magnetic-field \ac{SLAM}, in which the map \M is time-varying and only accurately models the field around the current location. As for all dead-reckoning techniques, the location error grows without bound. Notably, for magnetic-field dead reckoning to function in practice, measurements from an array of magnetometers are required.

The main properties of the three localization technologies are summarized in Tab.~\ref{tab:summary}.  Selected representative publications describing the implementation of the techniques for indoor and outdoor applications are also provided.

\newcolumntype{L}[1]{>{\raggedright\arraybackslash}p{#1}}
\begin{table*}[tb]
\caption{Summary of the properties of the magnetic-field localization technologies and selected representative publications implementing these techniques for indoor and outdoor localization.}
\label{tab:summary}
\centering
\renewcommand{\arraystretch}{1.1}
\begin{tabular}{
L{2.8cm}
L{4cm}
L{4.3cm}
L{1.8cm}
L{1.35cm}
L{1.35cm}
}
\toprule
\textbf{Magnetic-field localization technology} &
\textbf{Bounded localization error} &
\textbf{Map representation} &
\textbf{Requires sensor array} &
\textbf{Indoor application} &
\textbf{Outdoor application} \\
\midrule
Map matching & Yes &
Fixed known global-scale map &
No & \cite{solinSKR:2016,coulin2021tightly}
& \cite{canciani2016absolute,canciani:2017}
\\
\ac{SLAM} &
Yes, when revisiting mapped areas &
Inferred global-scale map &
No & \cite{kokS:2018,vallivaaraHKR:2010}
& \cite{leeC:2020}
\\
Dead reckoning &
No &
Inferred time-varying local-scale map &
Yes, in practice & \cite{huangHFPS:2024,Chesneau2016}
& \cite{Benedikt2025}
\\
\bottomrule
\end{tabular}
\end{table*}

The map \M used in these three localization techniques may be deterministic or stochastic. In the stochastic case, it is characterized by a probability density $p\bigl(\M\bigr)$ and can capture uncertainties about the map. In the subsequent discussion, the map is assumed to be stochastic, with the deterministic case regarded as a special instance corresponding to a degenerate density. 
    
From a Bayesian viewpoint, the listed localization techniques aim at computing one of the following posterior density functions, or an approximation thereof,
\begin{subequations}\label{eq:posterior densities}    
\begin{gather}
   p(x_t, \M \mid y_{1:t}, u_{1:t-1}),\label{eq:p(x,m|y)}\\
   p(x_t \mid y_{1:t}, u_{1:t-1}). \label{eq:p(x|y)}
\end{gather}
\end{subequations}
Here $u_{1:t-1}$ denotes measurements from potential auxiliary odometric sensors, such as inertial sensors or wheel encoders, used in the localization. Magnetic-field \ac{SLAM} and map matching aim at computing \eqref{eq:p(x,m|y)} and \eqref{eq:p(x|y)}, respectively. Magnetic-field dead reckoning aims to compute an approximation to \eqref{eq:p(x,m|y)} using a time-varying low-complexity map that is accurate in a small neighborhood of the current location.

Before continuing, two remarks should be made. First, the stated densities in \eqref{eq:posterior densities} correspond to a filtering inference setup, and corresponding densities for a smoothing inference setup can also be defined. Second, since the posterior density function in \eqref{eq:p(x|y)} can always be calculated from \eqref{eq:p(x,m|y)} via marginalizing over the map \M, the \ac{SLAM} localization technique is the most general localization technique. As will be shown later, the two other localization techniques can be viewed as special cases of \ac{SLAM}.

\section*{Mathematical Modeling}\label{sec:mathematical modeling}
\label{sec:modeling}
The calculation of the posterior densities in \eqref{eq:posterior densities} is commonly done using recursive application of Bayes' theorem and Chapman-Kolmogorov's equation. To that end, a system model for the dynamics of the localization system and the magnetic-field measurements of the form 
\begin{subequations}\label{eq:general Markov model}
\begin{gather}
    p(x_{t+1}\mid x_t,u_t)\label{eq:p(x|x)},\\
    p(y_t\mid x_t,\M)\label{eq:p(y|x)},\\
    p(x_0) \label{eq:p(x0)},\\
    p(\M) \label{eq:p(m)},
\end{gather}
\end{subequations}
is typically required. Here, the density in \eqref{eq:p(x|x)} models the dynamics of the localization system given the current state $x_t$ and the auxiliary input $u_t$. The likelihood in~\eqref{eq:p(y|x)} models the relationship between the state $x_t$, the map \M, and the magnetometer measurements $y_t$ at time instant $t$. Furthermore, the prior densities in \eqref{eq:p(x0)} and \eqref{eq:p(m)} capture knowledge about the initial state $x_0$ and the map \M. Next, commonly used magnetic-field models, magnetometer measurement models, and motion dynamic models will be presented.

\subsection*{Magnetic-field properties}\label{sec:magProperties}
The magnetic field \B has several important properties. Firstly, in linear media, such as air, it is additive and can be decomposed into the nominal Earth magnetic field and local spatial variations, a.k.a anomalies, caused by ferromagnetic materials in the Earth’s lithosphere and man-made structures. Secondly, the field is smooth away from sources; its magnitude approximately decays cubically with distance in the far field, and in non-conductive and non-magnetic media, the magnetic field propagates with negligible distortion and similar attenuation to that in air. Thirdly, \B is governed by Gauss's law for magnetism, which states that magnetic monopoles do not exist, and the field is therefore divergence-free. Finally, if the field is static, then in a domain with no magnetized materials or free currents, the Ampère–Maxwell law implies that the field is also curl-free. That is, inside this domain, it holds that
\begin{subequations}\label{eq:magnetic_field_properties}
\begin{align}
        \nabla_r \cdot \B(r) &= 0\quad \text{(divergence-free field)}, \label{eq:divergence_free}\\
      \nabla_r \times \B(r) &= 0\quad \text{(curl-free field)}. \label{eq:curl_free}
\end{align}
\end{subequations}
Therefore, in a simply connected domain, a static curl-free magnetic field can be represented as the gradient of a scalar potential. In this case, the divergence-free condition implies that the potential is harmonic. This, in turn, implies, for example, that for outdoor magnetic fields, the amplitude of horizontal spatial variations with wavelength $\lambda$ decays proportionally to $e^{-2\pi h/\lambda}$, where $h$ is the height above the Earth's surface~\cite{canciani2016absolute}. Hence, high-spatial-frequency variations are primarily observable close to the Earth's surface.


\subsection*{Magnetic-field modeling and maps}\label{sec:magModels}
A variety of mathematical models are used to create maps~\M. Examples of widely used models include discrete grid, polynomial, and \ac{GP} approximation-based function representations~\cite{viset:2024}. Most of these models can be abstractly viewed in terms of a basis function expansion of $L$ pre-determined basis functions $\phi_\ell(r)$, such as indicator function bases, polynomial bases, spectral bases, or radial function bases. That is, the map is modeled as 
\begin{equation}\label{eq:basisFunctions}
    \M(r) =\Phi^\Transp (r)w=\sum_{\ell=1}^{L}\phi^\Transp_\ell(r)\,w_\ell,
\end{equation}
where
\begin{align*}
    \Phi(r)&=\begin{bmatrix}
            \phi_1(r)\\
            \vdots\\
            \phi_{L}(r)
\end{bmatrix},&   w&=\begin{bmatrix} w_1 \\ \vdots \\ w_{L} \end{bmatrix}.
\end{align*}
Note that \eqref{eq:basisFunctions} is linear in the weights $w$, but typically nonlinear, both in the location $r$ and in the hyperparameters appearing in the basis functions~$\phi_\ell$.

The probability density $p\bigl(\M\bigr)$ of the map $\M$ is determined by the probability distribution $p(w)$ of the random weights~$w$. For instance, an approximate \ac{GP} model is obtained if the weights are normally distributed~\cite{solin2020}, where the choice of basis functions and weight covariance determines the corresponding covariance kernel of the process. Further, if the weights $w$ are drawn from a degenerate distribution, the model becomes deterministic.

\begin{figure*}[tb!]
\tcbset{raster equal height, raster columns=3, raster width=\textwidth}
\newlength{\figurewidth}
\setlength{\figurewidth}{.2\linewidth}
\newlength{\figureheight}
\setlength{\figureheight}{.2\linewidth}
\begin{tcolorbox}[width=\textwidth,colback={blue!10},title={\textbf{Examples of basis function expansions for map modeling}},colbacktitle=blue!20,coltitle=black]
\begin{tcbitemize}
Three commonly used basis function expansions for modeling magnetic-field maps \M. In the examples, the map (dashed black line) approximates the magnetic-field magnitude $\|B\|$ (solid black line) using four weighted basis functions $\phi_\ell(r)$. The colored lines show the weighted basis functions, i.e., $\phi_\ell^\Transp(r)w_\ell$.\\[1em]
  \tcbitem[title={\textbf{\makebox[0pt][l]{Grid-based map}}},colbacktitle=blue!20!white]
  \includegraphics[width=\columnwidth]{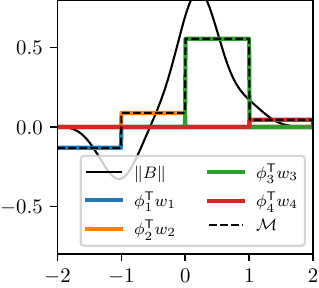}

  Map approximation using an indicator function basis
  \begin{equation*}
    \phi_\ell(r)
    =\begin{cases}
      1,& r\in\Omega_\ell,\\
      0,& \text{otherwise},
    \end{cases}
  \end{equation*}
  where $\Omega_\ell$ is the support of the $\ell$-th grid cell. Grid-based maps are commonly used in magnetic-field map matching and \ac{SLAM}, see, \eg,~\cite{canciani:2017,robertson:2013}. Due to the discontinuities, this type of map representation can neither encode global curl- nor divergence-free properties of the field. 

  \tcbitem[title={\textbf{Polynomial basis map}},colbacktitle=blue!20!white]
  \includegraphics[width=\columnwidth]{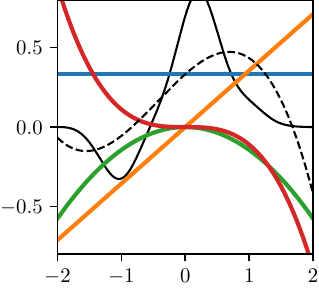}

  Map approximation using a polynomial basis 
  \begin{equation*}
      \phi_\ell(r)
      =(r-c_\ell)^\ell,
  \end{equation*} where $c_\ell$ is the center point. Polynomial models are commonly used to construct local-scale maps in magnetic-field dead-reckoning systems, see, \eg,~\cite{huangHFPS:2024,skog2018}. This type of map representation can encode both the curl- and divergence-free properties of the field.  

  \tcbitem[title={\textbf{Gaussian process map}},colbacktitle=blue!20!white]
  \includegraphics[width=\columnwidth]{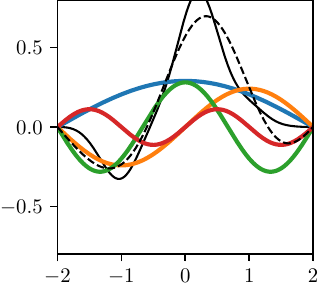}

Map approximation using a spectral basis 
\begin{equation*}
    \phi_\ell(r)
    = \cos(2\pi \ell r).
\end{equation*}
Spectral bases can be used to approximate \ac{GP} kernels and are commonly employed for global-scale magnetic-field maps in SLAM systems, see, \eg,~\cite{solin2020,kokSS:2024}. This type of map representation can encode both the curl- and divergence-free properties, but typically only the curl-free property is encoded.  
\end{tcbitemize}
\end{tcolorbox}
\end{figure*}

\subsubsection*{Incorporating curl and divergence properties}
There are several ways to incorporate the curl-free and divergence-free properties into the model of the field \B; when modeling the field magnitude $\lVert \B\rVert$, these properties are typically neglected. One approach to enforce the curl-free property is to let the map~\M represent a scalar potential, in which case the magnetic field is obtained as its negative gradient, i.e.,
\begin{equation}\label{eq:potential field map}
    \B(r) = -\nabla_r\,\M(r).
\end{equation}
The divergence-free property can be incorporated similarly by noting that the gradient is a linear operator~\cite{jidlingWWS:2017}. In this case, the map represents the vector potential. For polynomial function models, both curl- and divergence-free properties can be incorporated straightforwardly~\cite{skog2018}. For \ac{GP} models, 
this is less common; an exception is presented in~\cite{wahlstromKSG:2013}, where both properties are included. Typically, only the curl-free property is enforced in \ac{GP} models. 

Polynomial functions are ill-suited for large-scale magnetic-field modeling due to their poor extrapolation properties and numerical sensitivity at higher model orders. Consequently, polynomial functions are primarily used for local-scale magnetic-field models, such as those used in magnetic-field dead reckoning. For global-scale modeling, such as in magnetic-field map matching and \ac{SLAM}, models that incorporate only the curl-free property, or neither property, are typically adopted.

\subsubsection*{Other modeling techniques}
Nonlinear and nonparametric magnetic-field map representations have also been explored. Examples include curl- and divergence-free neural-network-based models and nonparametric grid-map representations based on stored magnetic-field measurements or features derived from them. However, as this article focuses on linear parametric map models of the form in \eqref{eq:basisFunctions}, such representations are not treated further.

\begin{figure*}[tb!]
\centering
\begin{tcolorbox}[width=\textwidth,colback={blue!10},title={\textbf{Magnetometer technologies}},colbacktitle=blue!20,coltitle=black] 
A wide range of magnetometer technologies based on different physical principles are available, with significant differences in measurement range, size, cost, and power consumption~\cite{Edelstein2007}. The technologies also differ in whether they directly measure the magnetic-field magnitude or have a specific sensitivity axis. The former are referred to as scalar or total-field magnetometers. In contrast, multiple sensors with a specific sensitivity axis can be combined to create a vector magnetometer that measures all components of the vector field. These properties largely determine the suitability of the different technologies for magnetic-field-based localization in indoor and outdoor environments. In the figure below, each box represents the typical root mean square noise level in a 1-hertz bandwidth and the maximum measurable field for three commonly used magnetometer technologies. Here, $T$ denotes Tesla. The indicated values correspond to compact magnetometer designs suitable for field applications and exclude higher-end technologies, such as \ac{SQUID} magnetometers, which require cooling for superconductivity.

\quad Magnetoresistive sensor technology is employed in contemporary chip-scale vector magnetometers, which are commonly used in indoor magnetic-field localization systems. The fluxgate technology can be used to construct vector magnetometers suitable for both indoor and outdoor localization systems. Fluxgate magnetometers are commonly used to construct magnetometer arrays for indoor magnetic-field dead reckoning (see, \eg,~\cite{huangHFPS:2024}), where the sensitivity of magnetoresistive sensors is insufficient to resolve small spatial variations in the magnetic field over short distances. Proton precision, Overhauser, and optically pumped magnetometers rely on quantum interference effects and offer very high sensitivity, which makes them well-suited for outdoor localization systems~\cite{lowrie2023earth}. Further, they can be designed to measure the field magnitude directly. This is advantageous outdoors, where spatial variations on the order of nano-Tesla must be resolved against a constant background field of about 50 micro-Tesla. Doing so with a vector magnetometer requires sensor orientation accuracy and sensitivity-axis orthogonality on the order of milli-degrees, which is practically challenging to achieve~\cite{Edelstein2007}.

\newcommand{\drawbox}[5]{%
  \def\halfheight{0.2}%
  \addplot [draw=#1, fill=#1, fill opacity=0.35, very thick]
    coordinates {(#2,#4-\halfheight) (#3,#4-\halfheight) (#3,#4+\halfheight) (#2,#4+\halfheight)} -- cycle;
  \node[font=\scriptsize\bfseries, text=white]
    at (axis cs:{sqrt(#2*#3)},#4) {#5};
  }
      
      \begin{tikzpicture}
        \definecolor{TradColor}{RGB}{31,119,180}   
        \definecolor{QuantColor}{RGB}{44,160,44}   
        \definecolor{RefColor}{RGB}{255,127,14}    

        \begin{axis}[
          width=\linewidth, height=0.62\linewidth,
          xmode=log, log basis x=10,
          xmin=1e-12, xmax=1,
          xlabel={Typical root mean square noise in a 1-hertz bandwidth and maximum measurable field (log scale)},
          xtick={1e-12,1e-9,1e-6,1e-3,1},
          xticklabels={pT,nT,$\mu$T,mT,T},
          ymajorticks=false, yminorgrids=false, ytick=\empty, yticklabels=\empty, axis y line*=none,
          ymin=0.0, ymax=2.7, y=0.9cm,
          axis x line*=bottom, grid=major, minor x tick num=9,
          tick label style={font=\small}, label style={font=\small},
          clip=false
          ]
          \def\EarthMin{25e-6}
          \def\EarthMax{65e-6}
          \addplot [draw=none, fill=RefColor!22]
          coordinates {(\EarthMin,0.0) (\EarthMax,0.0) (\EarthMax,2.6) (\EarthMin,2.6)} -- cycle;
          \node[font=\scriptsize, align=center, text=black, fill=white, fill opacity=0.8, text opacity=1, inner sep=1pt]
          at (axis cs:{sqrt(\EarthMin*\EarthMax)},2.15) {Earth field\\(25--65\,\textmu T)};

          \drawbox{QuantColor}{1e-12}{1e-4}{0.5}{Proton precision, Overhauser, and optically pumped}
          \drawbox{TradColor}{1e-10}{1e-3}{1}{Fluxgate}
          \drawbox{TradColor}{5e-9}{1e-2}{1.5}{Magnetoresistive}

          \node[anchor=south east, font=\scriptsize, align=right]
          at (rel axis cs:0.99,1.02) {{\color{QuantColor}\rule{8pt}{6pt}} Quantum\quad{\color{TradColor}\rule{8pt}{6pt}} Conventional\qquad\qquad\qquad};
        \end{axis}
\end{tikzpicture}

\end{tcolorbox}
\end{figure*}

\subsection*{Magnetometer measurement models}\label{sec:sensor models}
There exists a range of sensor technologies for measuring the magnetic fields. These can measure either the vector field~\B or its magnitude $\lVert \B \rVert$. Regardless of whether the vector field or the magnitude is measured, the measurements are typically assumed to be normally distributed and are modeled as
\begin{subequations}   
\begin{equation}
    p\bigl(y_t \mid x_t , \M\bigr) = \mathcal{N} \Bigl(y_t;h\bigl(x_t,\M\bigr), R_t \Bigr),
    \label{eq:measModel}
\end{equation}
where $h(\cdot)$ denotes a function that links the map \M to the expected mean of the measurements. Common functions are
\begin{equation}\label{eq:magfieldModel-vecpotmag}
h(x_t,\M)=
\begin{dcases*}
C^s_m(q_t)\M(r_t),            & if \M models the\\[-0.3em]
                    & vector field\\
-C^s_m(q_t)\nabla_r \M(r_t),   & if \M models the\\[-0.3em]
                    & potential field\\
\M(r_t),                & if \M models the\\[-0.3em]
                    & field magnitude
\end{dcases*}
\end{equation}
\end{subequations}
Here, $C^s_m(q)$ denotes a rotation matrix that transforms a vector from the map coordinate frame to the sensor coordinate frame. In magnetic-field \ac{SLAM} and map matching, the map coordinate frame is typically aligned with the global frame of reference, whereas in magnetic-field dead reckoning, the map is typically represented in the sensor frame. 
Furthermore, $R_t$ in \eqref{eq:measModel} denotes the covariance of the measurement noise, which is assumed to be uncorrelated across measurement time instants. As highlighted by \eqref{eq:magfieldModel-vecpotmag}, if a vector magnetometer is used, then the measurement depends on the orientation $q$ of the localization system.  

\begin{figure*}[tb!]
\centering
\begin{tcolorbox}[width=\textwidth,colback={blue!10},title={\textbf{Magnetometer measurement errors and calibration methods}},colbacktitle=blue!20,coltitle=black] 
Magnetometers are affected by intrinsic and extrinsic errors. Intrinsic errors arise from gain imperfections, offsets, non-perfectly orthogonal sensitivity axes in vector sensors, anisotropies in the sensing element, and thermal or quantization noise. Extrinsic errors arise from ferromagnetic materials in the sensor housing or the platform carrying the sensor. These are commonly divided into hard-iron effects, caused by permanent magnetic materials, and soft-iron effects, induced by ferromagnetic materials interacting with the Earth's magnetic field; the latter causes an orientation-dependent error. To first order, the combined effects of intrinsic and extrinsic errors can be described by the measurement model
\begin{equation*}
y_t = A_t(x_t)\,h\bigl(x_t,\M\bigr) + b_t(x_t) + e_t.
\end{equation*}
After calibrating $A_t(\cdot)$ and $b_t(\cdot)$, if the noise $e_t$ is modeled as normally distributed with covariance $R_t$, this model is equivalent to the one in~\eqref{eq:measModel}.

\quad Several methods exist for calibrating $A_t(\cdot)$ and $b_t(\cdot)$ from a sequence of measurements $y_{1:t}$ and states $x_{1:t}$. They differ in their assumptions about the magnetic field, the state trajectory $x_{1:t}$, and whether the magnetometer coordinate frame must be aligned with the coordinate frames of other sensors in the localization systems, such as inertial sensors. The choice of calibration method also depends on the availability of a calibration rig and on whether $A_t(\cdot)$ and $b_t(\cdot)$ are assumed time-invariant.

\quad A standard approach for calibrating vector magnetometers is ellipsoid fitting~\cite{Renaudin2010}. The method does not require $x_{1:t}$ to be known and exploits the fact that the Euclidean norm of an ideal sensor’s measurements is constant if \M is constant in the measurement volume. Since the Euclidean norm is invariant to rotation, the method cannot align the magnetometer coordinate frame with other reference frames. A review of calibration methods utilizing similar assumptions and techniques is given in~\cite{Papafotis2021}. Another standard approach for calibrating magnetometers mounted on aircraft and for compensating permanent, induced, and Eddy-current magnetic-field distortions is based on the Tolles–Lawson model~\cite{Han2017}. The model parameters are typically estimated from measurements collected during dedicated calibration maneuvers, i.e., $x_{1:t}$ is assumed known, using least-squares methods.

\quad Many magnetic-field localization systems also use auxiliary sensor measurements $u_{1:t}$, such as inertial sensors, when inferring the state $x_t$. Hence, it is often desirable to jointly calibrate the magnetometer and auxiliary sensors and to align their coordinate frames. To this end, several methods for joint magnetic-field and inertial-sensor calibration have been proposed; see, \eg,~\cite{kokS:2016}. Moreover, to avoid dedicated calibration procedures and to handle time-varying sensor errors, recent work has also explored methods for joint state inference and magnetometer calibration~\cite{sieblerLSH:2023,edridgeK:2026}.
\end{tcolorbox}
\end{figure*}

\subsection*{Models for the motion dynamics}
The motion dynamics of localization systems are defined by the platform on which they are mounted, and no universal motion dynamics model exists. That said, magnetometers are commonly used together with inertial sensors, wheel encoders, or other odometric sensors. The measurements $u_t$ from these sensors can be, but are commonly not, treated as measurements in the traditional sense by defining an additional likelihood function analogous to \eqref{eq:p(y|x)}. Instead, they are used as inputs to a deterministic dynamic model
\begin{equation}
x_{t+1} = f(x_t,u_t),
\end{equation}
which describes a dead-reckoning or inertial-navigation process; see~\cite{kokHS:2017} and~\cite{Farrell1998} for examples of such dynamic models.  

To take into account the measurement errors of the sensors, the motion dynamics of the localization system is, based upon a central limit theorem argument, commonly modeled as
\begin{equation}
p(x_{t+1}\mid x_t,u_t)
= \mathcal{N}\bigl(x_{t+1}; f(x_t,u_t), Q_t\bigr).
\end{equation}
Here, $Q_t$ is the process noise covariance matrix, which quantifies the effects of sensor errors, as well as numerical and approximation errors in the dead-reckoning or inertial-navigation process.

There are several reasons for treating the sensor measurements $u_t$ as inputs to the dynamic model and not as traditional measurements via a likelihood function~\cite{Farrell1998}. One reason is that inertial and odometric sensors provide information about the system's motion dynamics, rather than direct observations of the state $x_t$. Therefore, modeling these measurements via a likelihood function generally requires augmenting the state with additional state variables and associated dynamic models. This increases the complexity of both system modeling and inference. Another reason is that only the state propagation step of the inference process must be performed at the sample rate of the inertial or odometric sensors, which is often much higher than that of the magnetometers.

\section*{Statistical Inference for Map Learning}\label{sec:map learning}
Before examining the inference procedures used in magnetic-field-based localization techniques, it is helpful to review the basics about how a map can be inferred from a sequence of magnetometer measurements $y_{1:t}$ collected along a known state trajectory $x_{1:t}$. This is because map learning is a subtask of both magnetic-field \ac{SLAM} and magnetic-field dead reckoning. 

Inferring a map from measurements along a known trajectory $x_{1:t}$ is a relatively straightforward inference problem compared to the localization problem addressed later. Nevertheless, challenges remain, particularly regarding hyperparameter selection and the computational complexity of large-scale map inference. Much research has also been conducted on the related, more challenging problem of constructing maps from crowdsourced data when the state trajectory $x_{1:t}$ is not perfectly known. This type of map construction is out of the scope of this article, and the reader is referred to~\cite{Ouyang2022} and the references therein.

If a model of the form in~\eqref{eq:basisFunctions} is used to represent the map, inferring the map is equivalent to inferring the weight vector~$w$. To that end, assume that the measurements follow~\eqref{eq:measModel} and that the prior on the map $p(\M)$, or equivalently, the prior of the weight vector $w$, is
\begin{gather}
    p(\M)\equiv p(w) = \mathcal{N} \left(w; \mu_{w,0}, P_{ww,0} \right).
    \label{eq:p(w)}
\intertext{The posterior density of the weights $w$ at time~$t$ is then}
    p(w \mid y_{1:t}, x_{1:t}) = \mathcal{N} \left(w; \mu_{w,t}, P_{ww,t} \right),
\end{gather}
where the mean $\mu_{w,t}$ and covariance $P_{ww,t}$ can be computed using the stochastic recursive least-squares equations~\cite{Kailath2000}
\begin{subequations}
\begin{align}
    \mu_{w,t} &= \mu_{w,t-1} + K_t\,(y_t - H_t\,\mu_{w,t-1}), \\
    P_{ww,t} &= (\mathcal{I} - K_t\,H^\Transp_t)\,P_{ww,t-1},
\end{align}\label{eq:rls}%
where the gain $K_t$ is given by
\begin{equation}
        K_t = P_{ww,t-1}\,H^\Transp_t (H_t\,P_{ww,t-1}\,H^\Transp_t + R_t)^{-1},
\end{equation}
and
\begin{equation}
    H_t\triangleq \nabla_w\,h(x_t,\M). 
\end{equation}
\end{subequations}
Note, with $\M(r) =\Phi^\Transp (r)\,w$, then $h(x_t,\M)$ as defined in  \eqref{eq:magfieldModel-vecpotmag} is linear in the weights $w$.

The posterior density of the map at location $r'$ is finally given by
\begin{equation}
    p\bigl(\M(r') \mid y_{1:t}, x_{1:t}\bigr)
    = \mathcal{N}\bigl( \Phi^\Transp\!(r') \mu_{w,t}, \Phi^\Transp\!(r')  P_{ww,t} \Phi(r') \bigr).
\end{equation}
Next, the important task of selecting the hyperparameters in the model used to construct the map will be discussed.   

\subsection*{Hyperparameter selection}\label{sec:hyps}
To accurately infer the map $\mathcal{M}$, or equivalently the weights~$w$, the hyperparameters in the model \eqref{eq:basisFunctions} must be properly tuned as they determine the resolution and spatial variability of the inferred map. Important hyperparameters include the number of basis functions $L$, parameters that define the prior distribution of the weights $w$, and parameters that determine the spatial properties of the resulting map representation. 
Hyperparameters can be learned from data by maximizing the log-marginal likelihood~\cite{solin2018}, or more generally via hierarchical Bayesian inference. In practice, they are frequently set to physically motivated values, especially when the map is inferred recursively.

The number of basis functions $L$ required to accurately represent the magnetic field depends on both the desired spatial resolution and the spatial extent of the map. For local-scale models, $L$ can be selected using standard model-order selection criteria, such as the Akaike information criterion. For global-scale maps, including approximate \ac{GP}- and grid-based representations, $L$ is typically selected empirically to achieve the desired accuracy over the coverage area. 

The number of basis functions also directly impacts the computational complexity of the map inference. This is particularly critical for \ac{GP}-based models. While the computational complexity of exact \ac{GP} map inference scales cubically with the number of measurements, approximate \ac{GP} map inference scales linearly with the number of measurements and quadratically with the number of basis functions $L$. In these approximate models, the basis functions in \eqref{eq:basisFunctions} could, \eg, directly represent a subset of the measurements or a fixed number of inducing inputs \cite{quinoneroCandelaR:2005,viset:2024}, or could be a spectral basis \cite{solin2020,kokS:2018}. To further limit computational complexity, global-scale maps are sometimes represented using multiple intermediate-scale maps~\cite{kokS:2018} or approximated using basis functions with finite support~\cite{visetHK:2025}.

\section*{Statistical Inference for Localization}\label{sec:statistical inference techniques}
As previously stated, from a Bayesian perspective, the localization techniques
magnetic-field map matching, \ac{SLAM}, and dead reckoning aim to compute either of the posterior densities in
\eqref{eq:posterior densities}. Next, starting from the Bayesian filter recursions for magnetic-field \ac{SLAM}, it is shown that from a theoretical
viewpoint, magnetic-field map matching and dead reckoning can be viewed as special cases of magnetic-field \ac{SLAM}.
Thereafter, common implementation approaches and aspects of the three localization techniques are described.

Before continuing, recall that the map \M, as defined by the basis function
representation in \eqref{eq:basisFunctions}, is fully parameterized by the
weights $w$. With this representation, the posterior, likelihood, and prior can
be written equivalently as
$p(x_t, w \mid y_{1:t},u_{1:t-1})$, $p(y_t \mid x_t,w)$, and $p(w)$, respectively.
This notation is used hereafter.

The posterior densities for magnetic-field \ac{SLAM} in
\eqref{eq:p(x,m|y)} can, for $t>0$, be calculated using the system models in
\eqref{eq:general Markov model} and the Bayesian filter recursions~\cite{DurrantWhyte2006}
\begin{subequations}\label{eq:slam recursions}
    \begin{multline}
        p(x_t, w \mid y_{1:t},u_{1:t-1})\\ %
        \propto p(y_t \mid x_t,w)
        p(x_t, w \mid y_{1:t-1},u_{1:t-1}),
        \label{eq:UpdateSLAM}
    \end{multline}
    \vspace*{-5ex}
    \begin{multline}
        p(x_t, w \mid y_{1:t-1},u_{1:t-1})
        =\int p(x_t \mid x_{t-1}, u_{t-1})\\ %
        \cdot p(x_{t-1}, w \mid y_{1:t-1},u_{1:t-2})
        \,\mathrm{d}x_{t-1},
    \end{multline}
\end{subequations}
where the initial condition $p(x_0,w)=p(x_0)p(w)$ is given by the prior densities in
\eqref{eq:p(x0)} and \eqref{eq:p(w)}. The connection between the \ac{SLAM} inference problem and the earlier presented map inference problem can be seen by factorizing the \ac{SLAM} posterior density as follows
\begin{multline}\label{eq:factorization}
    p(x_{t}, w \mid y_{1:t},u_{1:t-1})\\
    =\int p(x_t,x_{1:t-1},w \mid y_{1:t},u_{1:t-1}) \mathrm{d} x_{1:t-1}\\
    =\int \underbrace{p(w \mid x_{1:t}, y_{1:t})}_{\text{Map inference}} \underbrace{p(x_{1:t} \mid y_{1:t},u_{1:t-1})}_{\text{State inference}} \mathrm{d} x_{1:t-1}.
\end{multline}
This factorization highlights that magnetic-field \ac{SLAM} consists of a coupled state- and map-inference problem. Given a trajectory posterior, the map posterior is obtained by conditioning on the entire state history. Marginalizing past states in the inferred trajectory recovers the SLAM posterior density. 

Magnetic-field map matching can be seen as a special case of \cref{eq:slam recursions}, where the posterior density in \eqref{eq:p(x|y)} can be calculated by, at every time instant $t$, marginalizing the posterior density with respect to the weights~$w$. This is not how magnetic-field map matching is implemented in practice, as it would require inferring the weights~$w$, which is computationally expensive. Instead, in magnetic-field map matching, it is assumed that prior knowledge of the map is sufficient for accurate localization. Therefore, the new information in the measurements $y_{1:t}$ and the trajectory $x_{1:t}$ about the map \M is neglected to make the approximation $p(w \mid x_{1:t}, y_{1:t})\approx p(w)$, which, if inserted in \eqref{eq:factorization}, results in the approximation 
\begin{equation}
    p(x_{t}, w \mid y_{1:t},u_{1:t-1})\approx p(w)p(x_{t}\mid y_{1:t},u_{1:t-1}).
\end{equation}
Next, inserting this approximation into \eqref{eq:slam recursions} and marginalizing with respect to the weights $w$, i.e., the map \M, yields the magnetic-field map-matching inference recursions 
\begin{subequations}\label{eq:map matching recursions}
    \begin{multline}
        p(x_t\mid y_{1:t},u_{1:t-1})\propto \underbrace{\int p(y_t \mid x_t, w)p(w) \mathrm{d} w}_{\text{ Marginal likelihood}}\\
        \cdot p(x_t\mid y_{1:t-1},u_{1:t-1}),\label{eq:map matching marginalization}
    \end{multline}
    \vspace*{-5ex}
    \begin{multline}
        p(x_t \mid y_{1:t-1},u_{1:t-1}) \approx\int p(x_t \mid x_{t-1}, u_{t-1})\\
        \cdot p(x_{t-1} \mid y_{1:t-1},u_{1:t-2}) \mathrm{d} x_{t-1}.
    \end{multline}
\end{subequations}
From \eqref{eq:map matching marginalization}, it can be seen that map matching is performed via the likelihood $p(y_t \mid x_t, w)$, where the uncertainty about the map is accounted for by marginalizing over $w$. If the map is a deterministic function, no marginalization is needed, further reducing computational complexity.    

Also, magnetic-field dead reckoning can be seen as a special case of magnetic-field \ac{SLAM}. In this case, the computational complexity of inferring the map $\M$ is reduced by instead using a time-varying local map $\M_t(r)=\Phi_t^\Transp(r)w_t$ that only describes the magnetic field (or its magnitude or potential) accurately in a local region around the current location. Since the map only needs to be accurate in a region around the current location, only a few basis function weights are needed to represent the map, saving memory and reducing computational complexity. By replacing the global-scale map $\M$ with the time-varying local-scale map $\M_t$, the magnetic-field \ac{SLAM} posterior density is approximated as
\begin{equation}
p(x_t, \M \mid y_{1},u_{1})
\approx
p(x_t, \M_t \mid y_{1},u_{1}).
\end{equation}
Further, the dynamics of the weights $w_t$ that parameterize the time-varying map $\M_t$ are modeled by the Markov transition density
\begin{equation}\label{eq:p(w_t+1|w_t)}
p(w_{t+1}\mid w_t, x_t,u_t),
\end{equation}
which captures the translation and rotation of the map with the state $x_t$ and auxiliary inputs $u_t$. Using the approximate posterior density, the parameterization $\M_t(r)=\Phi_t^\Transp(r)w_t$, and the dynamics for the time-varying map, the resulting Bayesian filter recursions for magnetic-field dead reckoning for $t>1$ become 
\begin{subequations}\label{eq:dead-reckoning recursions}
    \begin{multline}
            p(x_t, w_t \mid y_{1:t},u_{1:t-1})\\
            \propto p(y_t \mid x_t, w_t)
            p(x_t, w_t \mid y_{1:t-1},u_{1:t-1}),
    \end{multline}
    \vspace*{-5ex}
    \begin{multline}
        p(x_t, w_t \mid y_{1:t-1},u_{1:t-1})\\
        =\iint p(w_t \mid w_{t-1}, x_{t-1}, u_{t-1})
        p(x_t \mid x_{t-1}, u_{t-1})\\
        \cdot p(x_{t-1}, w_{t-1} \mid y_{1:t-1},u_{1:t-2})
        \,\mathrm{d}w_{t-1}\mathrm{d}x_{t-1}.
    \end{multline}
\end{subequations}
Here the initial condition $p(x_0,w_0)=p(x_0)p(w_0)$ is given by \eqref{eq:p(x0)} and the prior density of the local map $p(w_0)$. In the recursion, at each time step, the previous weights $w_{t-1}$ are marginalized. Although the marginalization over $w_{t-1}$ increases the per-step computational cost, this is outweighed by the substantial reduction in complexity obtained by representing the map with a low-dimensional weight vector~$w_t$ instead of the weight vector $w$ used in magnetic-field \ac{SLAM}. The price paid for the lower computational complexity is that no long-term global map is constructed. Hence, no loop closure can be performed after long exploration phases. As a result, the location error generally grows without bound. More details and views on the connection between magnetic-field \ac{SLAM} and dead reckoning are presented in~\cite{Skog2025}. 

To summarize, magnetic-field map matching and dead reckoning can be viewed as special cases of magnetic-field \ac{SLAM}, in which approximations are introduced to reduce computational complexity or memory requirements. Since all three localization techniques are special cases of magnetic-field \ac{SLAM}, they share several common inference and implementation challenges.

The inference process for all three localization techniques is nonlinear and must be solved using nonlinear filtering methods, such as the \ac{EKF}, \ac{UKF}, and \ac{PF}, or nonlinear optimization-based methods, such as \ac{FGO}. Further, if the orientation $q$ is a part of the state $x$, the fact that orientations belong to the Lie group of rotations $\mathrm{SO}(3)$ must be taken into account in the inference process. This holds regardless of whether the orientation is represented by a quaternion, rotation matrix, or Euler angles.

Further, in magnetic-field \ac{SLAM} and map matching, the posterior distributions are often multi-modal when the dynamic model $p(x_{t+1}\mid x_t,u_t)$ or the prior $p(x_0)$ is insufficiently informative. This behavior stems from the limited information content in a single magnetometer measurement; even in the absence of noise, a single measurement is generally insufficient to uniquely identify a location at a global scale. To that end, \ac{PF} techniques are frequently used in magnetic-field \ac{SLAM} and map matching, as they can represent multimodal posterior distributions. However, maintaining support for multiple modes over extended periods may require a large number of particles to avoid particle depletion and loss of low-probability hypotheses. Alternative methods for representing multimodal posteriors, such as particle smoothers and Gaussian sum filters, have only been explored to a limited extent in magnetic-field localization.

\begin{figure}[tb!]
\centering
\begin{tcolorbox}[width=\columnwidth,colback={blue!10},
title={\textbf{Inference of states including orientations}},colbacktitle=blue!20,coltitle=black] 
If the orientation is part of the state $x$, the inference problem involves a non-Euclidean space, since orientations belong to the Lie group of rotations $\mathrm{SO}(3)$. As a consequence, several operations used by standard nonlinear filtering techniques, such as the \ac{EKF} and \ac{UKF}, do not apply directly. For example, two orientations cannot, in general, be added. Therefore, these filters cannot be directly applied to states that include orientation components. In practice, this is typically handled by decomposing the orientation into a nominal part and an error part $\epsilon$, where the error $\epsilon$ is defined in a Euclidean space to which standard filtering techniques can be applied~\cite{Farrell1998}; the nominal state is maintained on the rotation manifold. This principle underlies commonly used invariant and error-state Kalman filtering techniques~\cite{barrau2016invariant,sola2017quaternion}.
\end{tcolorbox}
\end{figure}

\subsection*{Magnetic-field map matching}

Magnetic-field map matching is used both for indoor and outdoor applications. Indoors, 2D or full 3D spatial maps of the magnetic-field magnitude or vector field are typically constructed from one or more surveying campaigns. For outdoor applications, 2D maps of the magnetic-field magnitude are available, such as the Earth Magnetic Anomaly Grid\footnote{\url{https://www.ncei.noaa.gov/products/earth-magnetic-model-anomaly-grid-2}}. Methods that directly fit the Bayesian recursions from \eqref{eq:map matching recursions} are, \eg, \cite{solinSKR:2016,coulin2021tightly,canciani2016absolute}. 

An alternative and commonly used approach for magnetic-field map matching is sequence-based matching of magnetic measurements against existing maps; see, \eg, \cite{torresRMBH:2015,leeAH:2018}. These methods can be incorporated into the presented Bayesian framework by extending the measurement model in \eqref{eq:measModel} to represent sequences of measurements. That said, the maps used are typically nonparametric, in which measurements or features thereof are stored directly rather than represented by a finite-dimensional parametric model of the form in \eqref{eq:basisFunctions}.

\begin{figure*}
  \centering
  \includegraphics[width=\textwidth]{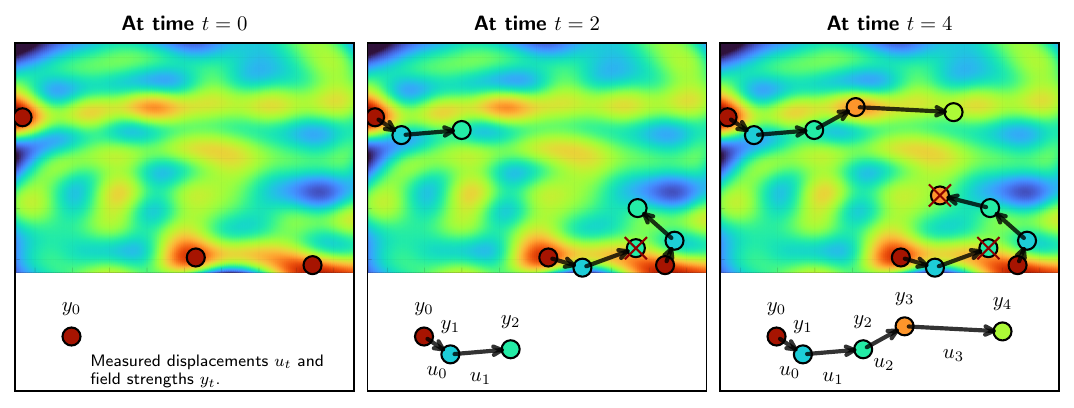}
  \caption{Illustration of localization using magnetic-field map matching. The inference algorithm is initialized with three equally probable locations at $t = 0$, and then alternates between comparing the magnetometer measurement $y_t$ with the map $\M$ and predicting the next state $x_{t+1}$ using the motion model $p(x_{t+1} \mid x_t, u_t)$. The graph below each image shows the dead-reckoning process (edges) and the available measurement information (nodes). By $t = 4$, only a single location hypothesis remains.}
  \label{fig:map matching}
\end{figure*}

The challenge with inference for magnetic-field map matching is the multimodality of the posterior due to the non-uniqueness of the magnetic field. This is illustrated in \cref{fig:map matching} (left), which shows that each measurement typically matches several locations in the map. Therefore, without an accurate prior on the initial location, a unique location can only be inferred over time, after sufficient movement of the localization system, see \cref{fig:map matching} (middle and right). Multimodality is also a concern when the dynamic model is inaccurate or in regions with limited spatial variations. 

The map-matching inference in \eqref{eq:map matching recursions} has been addressed using \ac{PF} and \ac{EKF}-based inference methods. The accuracy of these methods depends on the degree of magnetic-field variation and will therefore vary spatially and be highly environment-dependent; see~\cite{siebler2022bayesian}, where a Bayesian Cram\'er-Rao lower bound is derived. Estimation accuracy can degrade when the true posterior is multi-modal and an algorithm such as an \ac{EKF} is used, which can only track a single mode. Furthermore, magnetometer calibration is essential for magnetic-field map matching localization to avoid mismatches between the measurements during the localization process and the measurements used to create the map. To address this, recent work has explored simultaneous localization and calibration~\cite{sieblerLSH:2023}.

\subsection*{Magnetic-field \ac{SLAM}}\label{sec:inf-slam}
As of today, the main application of magnetic-field \ac{SLAM} is indoor localization, see \eg, \cite{kokS:2018,visetHK:2022,coulin2021tightly,vallivaaraHKR:2010}. That said, outdoor applications also exist; see, \eg, the aerial magnetic-field \ac{SLAM} system in~\cite{leeC:2020}. 

Most existing work on magnetic-field \ac{SLAM} directly fits the recursions \eqref{eq:slam recursions}. That said, there are notable exceptions, see, \eg, \cite{gaoR:2015,sieblerLSH:2024}, that match measurement sequences in a similar manner to magnetic-field map-matching approaches in~\cite{torresRMBH:2015}. After matching the measurement sequences, these \ac{SLAM} methods conduct an explicit loop closure. This is achieved by substituting the measurement likelihood in \eqref{eq:measModel} with a loop-closure likelihood that imposes the pseudo-measurement $r_t \approx r_{t-\tau}$, where $t-\tau$ denotes a past time instant, which, according to the measurement sequence matching procedure, corresponds to the same physical location as the current time instant $t$.

Although the magnetic-field \ac{SLAM} inference problem can be solved using standard nonlinear inference techniques, challenges arise from its high dimensionality. While $\dim(x_t)$ is typically on the order of ten, $\dim(w)$ is typically on the order of thousands or more. Hence, computational complexity is often a limiting factor in magnetic-field \ac{SLAM}. Unless the basis functions from \eqref{eq:basisFunctions} have finite support, there is no sparsity that can be exploited in the Bayesian recursions. Therefore, while \ac{FGO} is used for magnetic-field \ac{SLAM} methods that match measurement sequences and include explicit loop closures~\cite{sieblerLSH:2024}, existing work solving the recursive inference problem in \eqref{eq:slam recursions} has primarily focused on \acp{EKF}~\cite{visetHK:2022} and \ac{PF} methods~\cite{kokS:2018,kokSS:2024}. Note that, for particle methods, the state dimension appears prohibitive. However, since the inference of the basis function weights $w$ is a linear inference problem, see \eqref{eq:rls}, it is possible to use a Rao-Blackwellized \ac{PF} \cite{kokS:2018,kokSS:2024}. These methods only represent the state $x_t$ using particles, while every particle infers its own map $w$ using \eqref{eq:rls}.

The performance of magnetic-field \ac{SLAM} depends on the revisit frequency to previously mapped locations, the degree of magnetic-field variation, and the accuracy of the dynamic model. If previous locations are not revisited for long periods or when the dynamic model is inaccurate, the uncertainty in the inferred state grows, which limits the ability to exploit revisits to previously mapped areas. In addition, the measurement likelihood is directly affected by magnetometer calibration errors, which, in turn, degrade localization performance. To address this issue, recent work has investigated joint magnetometer calibration and magnetic-field \ac{SLAM} \cite{vallivaara2026saying,edridgeK:2026}.

\subsection*{Magnetic-field dead reckoning}
\begin{figure}[tb!]
    \centering
    \begin{tikzpicture}[
        node distance=.5cm,
        every node/.style={inner sep=0, outer sep=0},
        arrow/.style={thick, -{Latex[length=4mm,width=3mm]}}
    ]

    \node (img1) {\includegraphics[trim={0.1cm 5.7cm 1cm 0},clip,width=0.48\columnwidth]{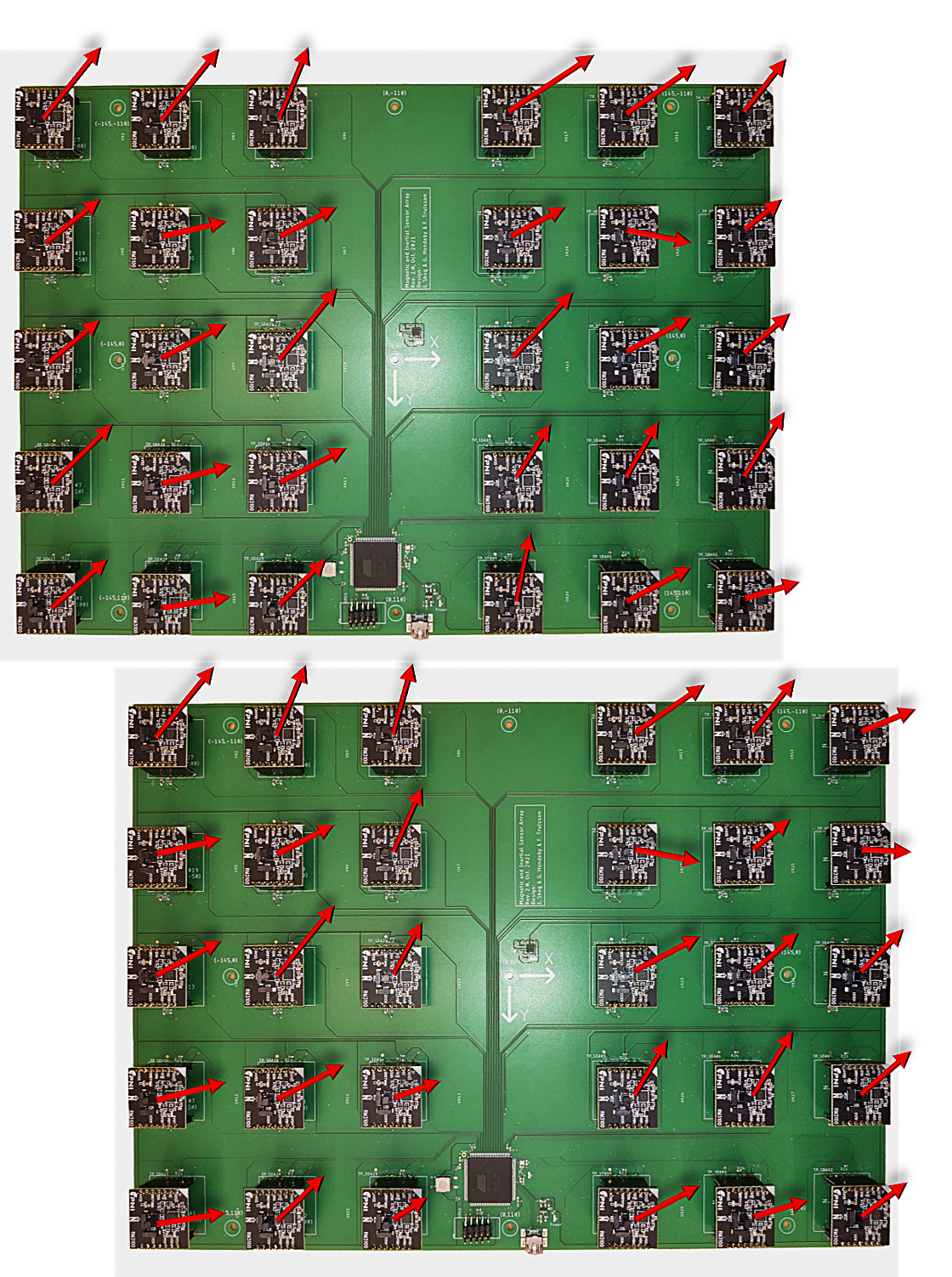}};
    \node[right=of img1] (img2) {\includegraphics[trim={1cm 0.2cm 0.2cm 5.8cm},clip,width=0.48\columnwidth]{figs/fig3.png}};

  \draw[arrow] 
  ([xshift=-1.3cm,yshift=.2cm]img1.north) 
  -- node[above, yshift=4pt, font=\small]{Change in location between $t$ and $t+1$} 
  ([xshift=1.3cm,yshift=.2cm]img2.north);

    \node[below=0.1cm of img1] {\small Location at $t$};
    \node[below=0.1cm of img2] {\small Location at $t+1$};
    \end{tikzpicture}
    \vspace{-1mm}
    \begin{tikzpicture}
\tikzset{
  sampleStem/.style={blue!70!black, thick},
  sampleMark/.style={blue!70!black, thick, fill=none}
}
\begin{axis}[
    width=1.1\columnwidth,
    height=0.55\columnwidth,
    axis lines=left,
      xlabel={$r$},
    ylabel={$\B(r)$},
    xlabel style={at={(axis description cs:0.95,-0.1)}, anchor=west},
   ylabel style={
    at={(axis description cs:0.1,0.85)},
    rotate=-90,
    anchor=south,
},
    xmin=0, xmax=10,
    ymin=0, ymax=4.2,
    xtick=\empty,
    ytick=\empty,
    clip=false, 
]

\def\magfield(#1){0.006*(#1)^3 - 0.1*(#1)^2 + 0.35*(#1) + 3}

\addplot[
    red,
    thick,
    domain=0:10,
    samples=200
]
{\magfield(x)};

\def\xA{1.1} \pgfmathsetmacro{\yA}{\magfield(\xA)}
\def\xB{2.0} \pgfmathsetmacro{\yB}{\magfield(\xB)}
\def\xC{2.9} \pgfmathsetmacro{\yC}{\magfield(\xC)}
\def\xD{7} \pgfmathsetmacro{\yD}{\magfield(\xD)}
\def\xE{7.9} \pgfmathsetmacro{\yE}{\magfield(\xE)}
\def\xF{8.8} \pgfmathsetmacro{\yF}{\magfield(\xF)}

\pgfplotsset{
  sampleStem/.style={blue!70!black, very thick},
  sampleMark/.style={blue!70!black, very thick, fill=white}
}

\draw[sampleStem] (axis cs:\xA,0) -- (axis cs:\xA,\yA);
\draw[sampleStem] (axis cs:\xB,0) -- (axis cs:\xB,\yB);
\draw[sampleStem] (axis cs:\xC,0) -- (axis cs:\xC,\yC);
\draw[sampleStem] (axis cs:\xD,0) -- (axis cs:\xD,\yD);
\draw[sampleStem] (axis cs:\xE,0) -- (axis cs:\xE,\yE);
\draw[sampleStem] (axis cs:\xF,0) -- (axis cs:\xF,\yF);

\draw[sampleMark] (axis cs:\xA,\yA) circle[radius=2.2pt];
\draw[sampleMark] (axis cs:\xB,\yB) circle[radius=2.2pt];
\draw[sampleMark] (axis cs:\xC,\yC) circle[radius=2.2pt];
\draw[sampleMark] (axis cs:\xD,\yD) circle[radius=2.2pt];
\draw[sampleMark] (axis cs:\xE,\yE) circle[radius=2.2pt];
\draw[sampleMark] (axis cs:\xF,\yF) circle[radius=2.2pt];

\draw[densely dashed] (axis cs:\xB,0.85) ellipse [x radius=1.3, y radius=0.45];
\draw[densely dashed] (axis cs:\xE,1) ellipse [x radius=1.3, y radius=0.45];

\draw[densely dashed]
  (axis cs:\xC+0.5,0.85) -- (axis cs:\xC+1.4,0.85)
  node[
    below,
    align=center
  ] {$\{y^{(n)}_{t}\}_{n=1}^{N}$};

\draw[densely dashed]
  (axis cs:\xD-0.5,1) -- (axis cs:\xD-1.3,1)
  node[above,align=center] {$\{y^{(n)}_{t+1}\}_{n=1}^{N}$};

\node[font=\small,rotate=-9.5] at (axis cs:5,3.3) {Fitted local model: $\M_t(r)$};

\draw[densely dashed] (axis cs:\xB,0) -- (axis cs:\xB,-0.55);
\draw[densely dashed] (axis cs:\xE,0) -- (axis cs:\xE,-0.55);
\draw[<->] (axis cs:\xB,-0.45) -- (axis cs:\xE,-0.45)
  node[midway,below] {Inferred displacement: $\Delta \hat{r}_t$};

\draw[<->] (axis cs:\xA,1.7) -- (axis cs:\xB,1.7) node[midway,above] {$d^{(1)}$};
\draw[<->] (axis cs:\xB,1.70) -- (axis cs:\xC,1.70) node[midway,above] {$d^{(2)}$};

\end{axis}
\end{tikzpicture}
    \caption{At top, an array with 30 vector magnetometers measuring the magnetic field at two consecutive locations. Below, a conceptual illustration in 1D of how the displacement $\Delta r_t$ can be inferred by fitting a local magnetic-field model $\M_t$ to the measurement from the two locations, and treating the displacement $\Delta r_t$ as an unknown parameter in the model.}\label{fig:array}
\end{figure}

As of today, most magnetic-field dead reckoning methods have been developed for pedestrian localization in indoor environments~\cite{huangHFPS:2024,Li2025,Chesneau2016}. Only a limited amount of work on magnetic-field dead reckoning for outdoor applications exists, with two examples being train localization~\cite{Benedikt2025} and autonomous underwater vehicle localization~\cite{skog2025underwater}. This may be attributed to the fact that indoor magnetic fields exhibit significant spatial variations over short length scales, whereas outdoor magnetic fields do not. Consequently, magnetic-field dead reckoning outdoors is more challenging and more sensitive to temporal variations in the magnetic field~\cite{skog2025underwater}. Notably, most magnetic-field dead-reckoning methods use measurements from an array of magnetometers, such as the one shown in Fig.~\ref{fig:array}, to make the likelihood sufficiently informative for reliable localization. That is, $p(y_t \mid w_t, x_t)$ in \eqref{eq:dead-reckoning recursions} is replaced by 
\begin{equation}
    \prod_{n=1}^{N} p(y^{(n)}_t \mid x_t, w_t, d^{(n)}),
\end{equation}
where $y^{(n)}_t$ denotes the measurement from the $n^{\text{th}}$ magnetometer located at $d^{(n)}$ in the array, and $N$ is the total number of magnetometers in the array. Note that $d^{(n)}$ is known and included to highlight that the magnetometer measurements are collected at different spatial locations.

Although recent state-of-the-art methods are based on the Bayesian recursions presented in \eqref{eq:dead-reckoning recursions}, the idea of magnetic-field dead reckoning was earlier approached via the differential equation~\cite{Vissiere2007a}
\begin{equation}\label{eq:diff B}
    \dot{\B}(r)=\B(r)\times \omega + \nabla_{r} \B(r)\,v .
\end{equation}
Equation~\eqref{eq:diff B} shows that, if the time derivative of the magnetic field $\dot{\B}(r)$, its spatial Jacobian $\nabla_{r}\B(r)$, and the angular rate $\omega$ are measured, the velocity $v$ can be inferred; all quantities are here expressed in the sensor coordinate frame. An estimate of the spatial Jacobian of the magnetic field can be obtained from measurements of an array of vector magnetometers, such as the planar array shown in Fig.~\ref{fig:array}. Notably, if the field is assumed to be curl- and divergence-free, a planar array of vector magnetometers is sufficient to recover the full three-dimensional spatial Jacobian of the field~\cite{skog2018}. By integrating the inferred velocity and the measured angular rate over time, magnetic-field dead reckoning can be performed. However, in practice, the inferred velocity is more commonly used to aid an inertial navigation system~\cite{Chesneau2016}.

More recently, the magnetic-field dead-reckoning problem has also been approached from a model-learning perspective~\cite{skog2018}. This approach is illustrated in Fig.~\ref{fig:array}, where a local time-varying magnetic-field model $\M_t$, parameterized by~$w_t$, is fitted to the array measurements from two consecutive time instants. The displacement $\Delta r_t$ and orientation change~$\Delta q_t$ of the array between $t$ and $t+1$ are treated as unknown parameters and inferred jointly with the parameters~$w_t$ of the magnetic-field model. That is, by defining the array coordinate frame at time $t$ as the reference frame, the posterior density
\begin{subequations}
\begin{multline}\label{eq:dispalcement learning}
    p(\Delta r_t, \Delta q_t, w_t \mid y^{(1)}_{t},\ldots,y^{(N)}_{t},y^{(1)}_{t+1},\ldots,y^{(N)}_{t+1})\\
    \shoveleft{\qquad\propto p(w_t)
    p(\Delta r_t, \Delta q_t)}\\
    \times\prod_{n=1}^{N}p\bigl(y^{(n)}_{t} \mid 0, w_t, d^{(n)} \bigr)\,p\bigl(y^{(n)}_{t+1} \mid \Delta x, w_t, d^{(n)}\bigr),
\end{multline}
is inferred. Here, the zero argument denotes the state at time $t$, which is all zero. Further, $p(\Delta r_t, \Delta q_t)$ represents prior information about the displacement and orientation change. Moreover, the state $\Delta x$ contains the displacement and orientation change and is defined as
\begin{equation}
      \Delta x=\begin{bmatrix}
        \Delta r_t \\
        \Delta q_t
    \end{bmatrix}.
\end{equation}
\end{subequations}
By repeating this inference process and accumulating the inferred displacements $\Delta r_t$ and orientation changes $\Delta q_t$ over time, dead reckoning can be performed. Notably, unlike the differential-equation-based approach, the model-learning approach can be used with both vector and total-field magnetometers.

Current state-of-the-art methods for magnetic-field dead reckoning implement the Bayesian filter recursions in \eqref{eq:dead-reckoning recursions} using an error-state Kalman filter, in which the posterior is approximated as normally distributed~\cite{huangHFPS:2024,Li2025}. A first-order curl- and divergence-free polynomial map $\M_t$ is typically used to represent the field. As these implementations use arrays holding many magnetometers, careful calibration to align the coordinate axes and the sensitivities of all magnetometers is required for the methods to work well.

As presented earlier, the Bayesian recursions for magnetic-field dead reckoning in \eqref{eq:dead-reckoning recursions} can be viewed as a special case of magnetic-field \ac{SLAM} with a local time-varying map $\M_t$ defined by the weights $w_t$. They can also be interpreted as a recursive extension of the model-learning approach in \eqref{eq:dispalcement learning} to magnetic-field dead reckoning, in which a dynamic model of the form \eqref{eq:p(w_t+1|w_t)} is introduced to describe the evolution of the local time-varying map $\M_t$. From a SLAM perspective, the fact that the local map $\M_t$ is accurate only in the vicinity of the current state and has finite memory due to the dynamics of the weights $w_t$ implies that no long-term magnetic-field memory can be exploited. Consequently, loop closure after long-term exploration phases, as in magnetic-field \ac{SLAM}, is not possible, and the localization error grows without bound. Conversely, when a magnetic-field \ac{SLAM} system is provided with measurements from a magnetometer array, it can, in theory, effectively behave as a magnetic-field dead-reckoning system during exploration of previously unmapped areas~\cite{Skog2025}; practically, the map in the \ac{SLAM} system is often too coarse to reliably infer small displacement and orientation changes. Hence, if inertial navigation is used as the backbone of the SLAM process, this can, in theory, substantially reduce the growth rate of the location error when new areas are mapped and extend the feasible exploration duration~\cite{huangHFPS:2024}; typically, the growth rate of the inertial navigation system localization error is reduced from cubic to linear in time.   
 
Currently, no performance bounds for magnetic-field dead reckoning exist. However, Cram\'{e}r-Rao lower bounds on the displacement (velocity) using the model-learning approach were derived in~\cite{skog2018} and~\cite{Benedikt2025}, assuming polynomial and Gaussian-process maps, respectively. The bound in~\cite{skog2018} showed that, if the magnetic field can be locally approximated by a first-order polynomial, the estimation accuracy is inversely proportional to the slope-to-noise ratio and the square of the array length. Recently, a similar Cram\'{e}r-Rao lower bound that also considers the temporal variations in the field was derived. The bound shows that for outdoor magnetic-field dead reckoning at low speed, such as in underwater applications, temporal variations significantly affect accuracy, often surpassing the effect of sensor noise~\cite{skog2025underwater}.

\section*{Challenges and Outlook}
While research to date has demonstrated the feasibility of using spatial variations in the magnetic field for localization, many signal-processing challenges remain before these localization systems can be commercially deployed at scale. Moreover, advances in quantum magnetometer sensor technology are enabling the manufacture of highly sensitive, small-form-factor, low-power sensors, with noise levels on the order of pT/$\sqrt{\text{Hz}}$. Currently, the noise levels of these sensors are determined by the readout process rather than by the underlying quantum physics, whose noise level is at least one order of magnitude lower~\cite{degen2017quantum}. Hence, as readout technology improves, sensor noise will, in an increasing number of applications, no longer be the dominant source of error. Instead, errors due to model imperfections, unmodeled platform disturbances, and inference approximations increasingly dominate, which calls for more advanced signal-processing methods for modeling, calibration, and inference.

Currently, there is a lack of theoretical and analytically interpretable bounds on the achievable localization accuracy for magnetic-field SLAM and magnetic-field dead reckoning; such bounds exist for magnetic-field map matching, see, \eg,~\cite{siebler2022bayesian}. In addition, observability analyses characterizing the conditions under which system states, calibration parameters, and magnetic-field map parameters can be inferred are largely missing. Consequently, it remains difficult to analyze trade-offs in sensor and system designs and configurations, as well as the role of trajectory and motion excitation in the inference process. Such bounds and observability analysis results are also required when magnetic-field-based localization is to be combined with path and motion planning, where localization performance must be predicted as a function of the planned path and motion.

Scalability and computational efficiency remain key bottlenecks, especially when aiming for real-time estimation. Current map and state inference techniques for magnetic-field \ac{SLAM} are computationally demanding, and scaling these to large environments is still a challenge~\cite{visetHK:2025,kokS:2018}. Here, recent advances in sparse \ac{GP} approximations, distributed \ac{GP} regression, and scalable Bayesian inference methods may offer solutions. 

All three magnetic-field localization techniques can be formulated using 1D, 2D, and 3D maps. The dimensionality directly affects the memory requirements and computational complexity of the inference process. Beyond this, the dimensionality also affects the ability to reduce uncertainty in the inferred state $x_t$ and map $\M$, since higher dimensionality implies that the information provided by the measurements must be shared across more parameters. This is particularly pronounced in magnetic-field \ac{SLAM}, where a global-scale map is inferred jointly with the state trajectory. Consequently, many experimental evaluations of 2D and 3D magnetic-field \ac{SLAM} systems rely on trajectories with substantial self-overlap, effectively reducing the dimensionality of the inference problem, at least temporarily. How well these systems function across trajectories with varying degrees of self-overlap remains an open question.

Several publicly available datasets and open-source implementations for magnetic-field localization exist, see \eg,~\cite{huangHFPS:2024,torresRMBH:2015,solin2018,visetHK:2022,kokSS:2024}. Notably, these publications are almost exclusively focused on indoor localization. For outdoor applications, global and regional magnetic field databases, such as the Earth Magnetic Anomaly Grid, provide magnetic-field maps that may be used for map-matching. That said, publicly available datasets that combine magnetic measurements with ground-truth trajectories for outdoor, large-scale scenarios remain scarce, hindering benchmarking and reproducibility. One example of such a dataset is the DAF-MIT AIA airborne magnetic navigation dataset~\cite{muradoglu2020}.

Map construction and maintenance remain challenging in many application domains. National institutes frequently conduct surveys to create maps of the Earth's magnetic field, which may be used for map-matching in aerospace and on-surface applications~\cite{Goldenberg2006}. However, with current methods, the Earth’s magnetic field can only be reliably extrapolated upward, not downward, relative to the height at which the field was mapped. Hence, when aerial survey maps are used for marine applications, the finite spatial bandwidth of the magnetic-field map limits which spatial-frequency components of the measurements can be reliably matched to the map. This limits the achievable localization accuracy and requires speed-dependent low-pass filtering of the measurements.

\begin{figure*}[tb!]
\centering
\begin{tcolorbox}[width=\textwidth,colback={blue!10},title={\textbf{Looking beyond the presented models and inference techniques}},colbacktitle=blue!20,coltitle=black] 
The presented map and sensor models have primarily been linear parametric models based on physical knowledge of the underlying process or system, and designed to at least partially comply with its physics. Further, the presented inference techniques have been based on recursive Bayesian inference. 

\quad Classical machine-learning methods, such as $k$-nearest neighbors and support vector machines, have also been widely used for magnetic-field map matching~\cite{Ouyang2022}. Recent research has also explored neural-network-based models and inference methods.

\quad For magnetic-field modeling, physics-informed neural-network models that enforce curl- and divergence-free magnetic fields have been explored, see, \eg,~\cite{magadA2026}. Furthermore, neural networks have been used to extend classical calibration methods, such as Tolles--Lawson-type models, to compensate for platform-induced magnetic interference~\cite{Xu2024interference,hagerA2026}. This is particularly important when the platform's magnetic signature varies over time, \eg, due to payload changes, or when sensors are mounted in close proximity to the platform's main body. In addition, neural networks have been used for end-to-end inference, where the network directly infers the location or computes similarity scores from sequences of magnetic-field measurements without an explicit intermediate map representation, see e.g.,~\cite{ashrafI2020,leeAH:2018}.

\quad Compared with the presented linear parametric models, these neural network models provide greater modeling flexibility and can represent more complex magnetic-field, sensor, and platform disturbance behaviors. However, this flexibility typically comes at the cost of a more computationally demanding inference procedure and a need for more measurements to accurately infer the model parameters. Further, end-to-end inference methods, as of today, seldom compute the full posterior density, making rigorous quantification of the uncertainty in the inferred location challenging. 
\end{tcolorbox}
\end{figure*}

The map maintenance challenges are further compounded by temporal variations in the ambient magnetic field. In outdoor magnetic-field localization of high-velocity platforms, the spatial variations appear as a high-frequency component that can be separated from low-frequency temporal variations~\cite{canciani2016absolute}. However, for slow-moving platforms, the spatial and temporal field components cannot be separated, and correction data from a reference sensor is typically required. Alternatively, for magnetic-field map-matching localization, gradient-based map matching using magnetometer arrays may be used~\cite{Xu2024}. The use of magnetic-field gradients in magnetic-field \ac{SLAM} remains unexplored.

Current methods for indoor magnetic-field localization that exploit spatial variations in the magnetic field typically neglect temporal field variations. While temporal variations in the Earth’s magnetic field can often be neglected indoors, changes in building layouts, furniture, or equipment may cause abrupt changes in the spatial field, leading to erroneous maps. How to detect, model, and adapt to such changes remains largely unexplored~\cite{solin2018}.

In magnetic-field \ac{SLAM}, access to accurate dead reckoning or inertial navigation is fundamental to enable extended exploration phases in which new areas are mapped. Contemporary consumer-grade inertial sensors generally lack the accuracy required to enable practically usable magnetic-field SLAM systems for indoor localization. Recent research indicates that magnetic-field SLAM using arrays of magnetometers, thereby inheriting properties of magnetic-field dead reckoning, may offer a potential solution~\cite{Skog2025}.

Another system-level challenge, extending beyond magnetic-field-based localization, is the need for methods to monitor the consistency and integrity of the inferred localization information. To date, most research on magnetic-field-based localization has focused on maximizing the accuracy of inferred point estimates of the location. However, consistency and integrity monitoring are fundamental if the localization information is to be used in control and decision-making systems and in safety-critical applications.

Finally, the signal-processing principles, methods, and algorithms developed for magnetic-field localization have the potential to transfer to other physical fields. For example, gravity-based and terrain-based navigation share many characteristics with magnetic-field localization and could benefit from similar modeling and inference techniques.

\section*{Acknowledgment}
This work has been partially funded by the Swedish Research Council project 2020-04253 \emph{Tensor-field based localization} and Vinnova, Sweden’s Innovation Agency, and Swedish Armed Forces, project 2025-03855 \emph{Electromagnetic navigation for smaller unmanned underwater vehicles}. This work was also partially performed within the Competence Center SEDDIT (Sensor Informatics and Decision making for the DIgital Transformation), supported by Sweden’s Innovation Agency within the research and innovation program Advanced digitalization. This work was also supported by the Sensor AI Lab through the AI Labs Program of the Delft University of Technology.

\bibliographystyle{IEEEtran}
\bibliography{IEEEfull,refs}
\section*{Authors}
\begin{IEEEbiographynophoto}{Isaac Skog} (\url{skog@kth.se}) received his MSc degree in Electrical Engineering from KTH Royal Institute of Technology, Stockholm, Sweden, in 2005. In 2010, he received a PhD degree in Signal Processing from KTH Royal Institute of Technology with a thesis on low-cost navigation systems. Currently, he is an associate professor in communication systems at KTH and a senior researcher at the Swedish Defence Research Agency (FOI) in Stockholm, Sweden. He is an editor of the IEEE Journal of Indoor and Seamless Positioning and Navigation. His research interest is in applied signal processing for localization and tracking. 
\end{IEEEbiographynophoto}

\begin{IEEEbiographynophoto}{Manon Kok} (\url{m.kok-1@tudelft.nl}) received the M.Sc.\ degrees in Applied Physics and in Philosophy of Science, Technology and Society, both from the University of Twente, the Netherlands, and the Ph.D. degree in Automatic Control from Link\"oping University, Sweden. Previously, she was a Research Associate at the University of Cambridge, U.K. She is currently an Associate Professor at the Delft University of Technology, the Netherlands. She is Associate Editor for the IEEE Trans. on Aerospace and Electronic Systems, part of the Steering Committee of the IEEE Journal of Indoor and Seamless Positioning and Navigation, and serves on the ISIF Board of Directors. 
\end{IEEEbiographynophoto}

\begin{IEEEbiographynophoto}{Christophe Prieur} (\url{christophe.prieur@gipsa-lab.fr}) is a senior researcher of the CNRS, Grenoble, France. He was the Program Chair of the 9th IFAC Symposium on Nonlinear Control Systems, the 14th European Control Conference and the 61st IEEE Conference on Decision and Control. He has served as associate editor and senior editor of several journals, and he is currently an editor of the IMA Journal of Mathematical Control and Information. He was awarded the
CNRS Silver Medal. He is an IMA Fellow and an IEEE Fellow. His current research interests include navigation, object tracking, and fluid/thermal dynamics control.
\end{IEEEbiographynophoto}

\begin{IEEEbiographynophoto}{Gustaf Hendeby} (\url{gustaf.hendeby@liu.se})
is an Associate Professor in Automatic Control at Linköping University, Sweden.
He received his M.Sc.\ in Physics and Engineering (2002) and Ph.D.\ in Automatic Control (2008) from Linköping University, and held positions at DFKI and FOI.
He is Editor for IEEE Transactions on Aerospace and Electronic Systems and Associate Editor-in-Chief of Journal of Advances in Information Fusion.
He leads the WASP Localization and Navigation Cluster, has chaired multiple FUSION conferences, and serves on the ISIF Board of Directors as Vice President.
His research focuses on sensor fusion and signal processing, including nonlinear estimation, target tracking, and SLAM.
\end{IEEEbiographynophoto}

\end{document}